\documentclass[epj]{svjour_ag}
\usepackage{graphicx}
\begin{document}                                                        
\journalname{Physica E, in press}
\title{About universality of lifetime statistics in  
       quantum chaotic scattering}
\author{M. Gl\"uck, A. R. Kolovsky\cite{byline} and H. J. Korsch}
\institute{Fachbereich Physik, Universit\"at Kaiserslautern, D-67653 
         Kaiserslautern, Germany}
\date{\today }


\abstract{
The statistics of the resonance widths and the behavior of the
survival probability is studied in a particular model of quantum
chaotic scattering (a particle in a periodic potential subject to
static and time-periodic forces)
introduced earlier in Ref.~\cite{PRA1,PRL1}.
The coarse-grained distribution of the resonance widths
is shown to be in good agreement with the prediction of Random
Matrix Theory (RMT). The behavior of the survival probability
shows, however, some deviation from RMT.\\[0.2cm]
}

\titlerunning{Lifetime statistics in  
       quantum chaotic scattering}
\authorrunning{M.~Gl\"uck, A.~R.~Kolovsky and H. J. Korsch }
\maketitle

{\bf 1.}
The abstract Random Matrix Theory (RMT) is known as a 
powerful tool for analyzing complex quantum systems.
During the last two decades the predictions of the {\em hermitian}
RMT (which is supposed to describe the properties of a closed
system) were checked for a large number of physical models
and an understanding of the conditions of applicability
of RMT was reached. The situation is different, however, for
{\em nonhermitian} RMT, which is aimed to describe the spectral
properties of open systems. Here we have quite a few physical models
which can serve a suitable test for a nonhermitian RMT.
Up to our knowledge the following systems are mainly under discussion: 
chaotic 2D billiards with attached leads \cite{billiard}; 
the kicked rotor with an absorbing boundary condition \cite{rotor}; 
simplified models of a dissociating molecule \cite{molecule};
scattering on graphs \cite{graph};
and the Bloch particle affected by static and time periodic forces
(1D model of a crystal electron in dc and ac fields) \cite{PRA1,PRL1,PRE2}. 
The latter model has a number of nice features,
which distinguishes it among the other systems. First, it is simple 
for numerical analysis. Second, it always realizes the so-called 
case of perfect coupling (where the predictions of the hermitian
and nonhermitian RMT are most different). Third, as a physical model
it can be and has been studied under laboratory conditions \cite{raizen}.
This present paper continues our study of the Bloch particle
in ac and dc fields in relation to RMT. 
In particular we discuss here the decay of the probability 
in the system  (probability leakage), which is the simplest quantity
measured in the laboratory experiment.

\smallskip 
{\bf 2.}
We briefly recall some of the results of our preceding papers
\cite{PRA1,PRL1,PRE2}.
After an appropriate rescaling the Hamiltonian of the system of interest
can be presented in the dimensionless form
\begin{equation}
\label{1}
\widehat{H}=\frac{\hat{p}^2}{2}+\cos x+Fx+F_\omega x\cos(\omega t) \;.
\end{equation}
The parameters of the system (\ref{1}) are the amplitude of the
static force $F$, the amplitude $F_\omega$ and the frequency $\omega$
of the time-periodic force, and the scaled Planck constant $\hbar$
which enters the momentum operator $\hat{p}=-i\hbar{\rm d}/{\rm d}x$.
Another, unitary equivalent, form of the Hamiltonian (\ref{1}) reads as
\begin{equation}
\label{2}
\widehat{H}=\frac{\hat{p}^2}{2}+\cos[x-\epsilon\cos(\omega t)]+Fx 
\;,\;  \epsilon=\frac{F_\omega}{\omega^2} \;.
\end{equation}

\begin{figure}[h]
\begin{center}
\includegraphics[height=11.5cm,clip]{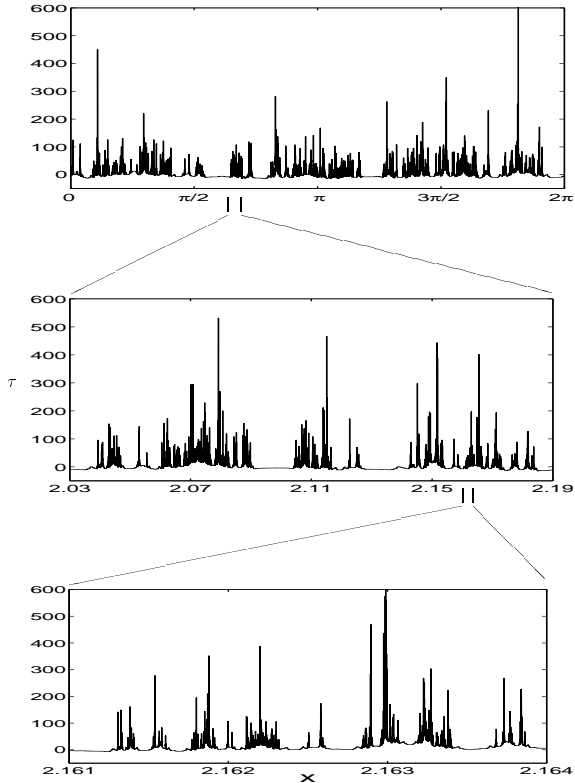}
\end{center}
\caption{Fractal structure of the classical delay time (\protect\ref{3})
as a function of the initial coordinate. The system parameters are
$F=0.3$, $\epsilon=1.5$, and $\omega=10/6$.} \label{fig1}
\end{figure}
When some condition (based on Chirikov's overlap criterion) on
$\epsilon$ is satisfied, the classical dynamics of the system (\ref{2})
is an example of chaotic scattering. In fact, one of the main 
characteristics of the classical chaotic scattering is the delay or 
dwell time
\begin{equation}
\label{3}
\tau=\lim_{p_0\rightarrow\infty}[\tau(p_0\rightarrow -p_0)-2p_0/F] \;.
\end{equation}
(Here $\tau(p_0\rightarrow -p_0)$ is the time taken by the particle to
change its initial momentum $p_0$ to the opposite one.) Figure \ref{fig1}
shows the delay time (\ref{3}) as a function of the initial coordinate $x$.
The fractal behavior is typical for chaotic scattering.

We proceed with the quantum case. It was shown in Ref.~\cite{PRL1,PRE2}
that the dynamics and spectral properties of the system (\ref{2})
depend crucially on the condition of commensurability between
the so-called Bloch period $T_B=\hbar/F$ and the period 
$T_\omega=2\pi/\omega$ of the driving force,
\begin{equation}
\label{4}
\frac{T_B}{T_\omega}=\frac{r}{q} \;.
\end{equation}
Providing the condition (\ref{4}) is satisfied, the complex quasi\-energy
spectrum of the system (the resonances) is given by the eigenvalues of
the following nonunitary matrix
\begin{equation}
\label{5}
U_{sys}=\left(
\begin{array}{cc}
0_{M\times N}&0_{M\times M}\\ W_{N\times N}&0_{N\times M}
\end{array}
\right)   \;.
\end{equation}
In Eq.~(\ref{5}) $W_{N\times N}$ is the unitary matrix with the
coefficients 
\begin{equation}
\label{6}
W_{n',n}=\langle n'|\exp(-ikx)\widehat{W}\exp(ikx)|n\rangle \;,
\end{equation}
\begin{displaymath}
\widehat{W}=\widehat{\exp}\left\{-\frac{i}{\hbar}\int_0^T
\left[\frac{(\hat{p}-Ft+\hbar k)^2}{2}+\widehat{V}\right]{\rm d}t\right\} \;,
\end{displaymath}
\begin{displaymath}
\widehat{V}=\cos[x-\epsilon\cos(\omega t)] \;,
\end{displaymath}
and $0_{M\times N}$, $0_{M\times M}$, $0_{N\times M}$ are blocks
of zeros. The nonunitary matrix (\ref{5}) can be thought of as the truncated
[to the size $(N+M)\times(N+M)$] unitary matrix of the system evolution
operator over the common period $T=qT_B=rT_\omega$. 
Then the parameter $M$, which plays the role of the number of
channels, is identical with the integer $q$ in condition (\ref{4}). 
The additional parameter $N$ measures the number of states supported 
by the chaotic component of the classical phase space. (Unimportant for 
our present aim is the quasimomentum $k$ which can take any
value in the interval $-\pi/q\le k<\pi/q$.)

The main conjecture made in Ref.~\cite{PRE2} is that the spectral 
statistics of the system (\ref{2}) [i.e., the eigenvalues statistics 
of the matrix $U_{sys}$] is the same as the statistics of the
eigenvalues of a random matrix $U_{ran}$ of the structure 
(\ref{5}) but with the matrix $W_{N\times N}$ substituted by a member
of the Circular Unitary Ensemble (CUE)
\begin{equation}
\label{7}
W_{N\times N}\rightarrow A_{N\times N} 
\;,\quad A_{N\times N} \in {\rm CUE} \;.
\end{equation}
In what follows we examine this conjecture in more detail.

\begin{figure}
\begin{center}
\includegraphics[height=6.0cm,clip]{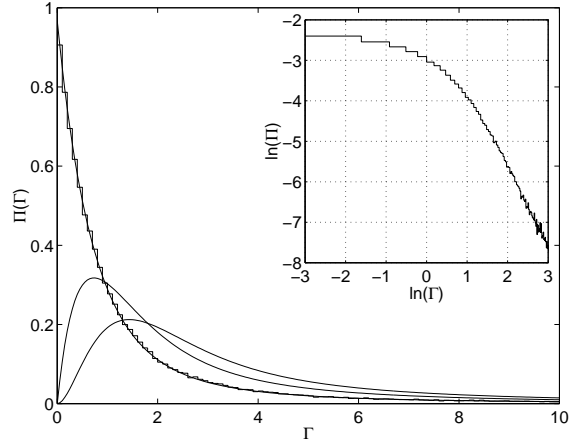}
\end{center}
\caption{Distribution of the scaled resonance width $\Gamma_s$
of the nonunitary random matrix $U_{ran}$. Parameters are $N=41$ and
$M=1$. The statistical ensemble involves 5000 matrices.
The smooth curves are the distribution (\protect\ref{8}) for $M=1$, $2$, 
$3$, respectively.} \label{fig2}
\end{figure}

\smallskip   
{\bf 3.}
First we discuss the spectral statistics of the random matrix $U_{ran}$.
There is strong numerical evidence that the statistics of the
eigenvalues $\exp(-i{\cal E})=\exp(-iE-\Gamma/2)$ of the matrix $U_{ran}$
is given by the universal distribution derived in Ref.~\cite{fyodorov}.
(An analytical proof of this result is still an open problem. 
  \footnote{After the paper was submitted we learned that this 
   result has been proved in Ref.~\cite{zyczkowski}.})
In particular, the distribution of the scaled resonance width
$\Gamma_s=\pi\Gamma/\Delta$ ($\Delta=2\pi/N$ is the mean level spacing) 
obeys the equation 
\begin{equation}
\label{8}
\Pi_M(\Gamma_s)=\frac{(-1)^M \Gamma_s^{M-1}}{(M-1)!}
\frac{d^M}{d\Gamma_s^M}
\left(e^{-\Gamma_s}\frac{\sinh \Gamma_s}{\Gamma_s}\right) \;. 
\end{equation}
%
As an example Fig.~\ref{fig2} (adopted from paper \cite{PRE2})
shows the histogram for the distribution of $\Gamma_s$ for $N=41$ 
and $M=1$. We note that $\Pi_M(\Gamma_s)\approx M/\Gamma_s^2$ for
$\Gamma_s\gg 1$ and, thus, the notion of mean resonance width is not 
defined.

Let us discuss now the decay of the probability $P(t)$, which is
given by the following equations
\begin{equation}
\label{9}
P(t)=|{\bf c}_t|^2 \;,\quad {\bf c}_{t+1}=U_{ran}{\bf c}_t
\;,\quad |{\bf c}_0|=1 \;.
\end{equation}
It is obvious that the dynamics of $P(t)$ is determined by
the spectrum of the system. Thus the behavior of $P(t)$ suggests
an additional test of the eigenvalue statistics \cite{alhassid}. 
The main advantage of studying $P(t)$ is that this quantity is more 
easily measured in the laboratory experiments \cite{fractal}.

\begin{figure}
\begin{center}
\includegraphics[height=7.0cm,clip]{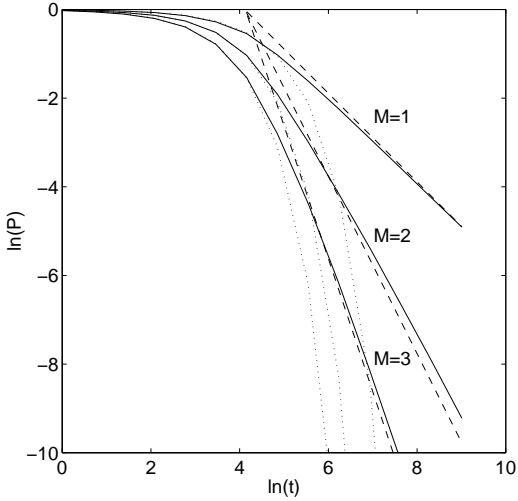}
\end{center}
\caption{Decay of the probability for the random matrix model (\protect\ref{9})
for $N=121$ and $M=1,2,3$. (Statistical ensemble involves 100 matrices).
The dotted and dashed lines correspond to the asymptotic 
(\protect\ref{10}).} \label{fig3}
\end{figure}

The problem of probability decay was considered (although
within a different RMT model) in paper \cite{sokolov}. It was proved
there that the function $P(t)$ has an exponential short-term
and an algebraic long-term asymptotic. Adopting these
results to the present model (\ref{9}) we obtain
\begin{equation}
\label{10}
P(t)=\left\{
\begin{array}{rr}
\exp\left(-\frac{M}{N}t\right)&,\quad t\ll t^*\\ 
\left(\frac{2t}{N}\right)^{-M}&,\quad t\gg t^*
\end{array}
\right.   \;,
\end{equation}
where $t^*\sim N/2=\pi/\Delta$ is of order of the Heisenberg time.
The results of a numerical simulation of the dynamics of $P(t)$ depicted 
in Fig.~\ref{fig3} well support the analytical expression (\ref{10}).

It is interesting to note that Eq.~(\ref{10}) can be obtained by using
rather simple arguments. In fact, expanding the initial vector
${\bf c}_0$ over the set of eigenvectors of the matrix $U_{ran}$
and following Ref.~\cite{sokolov} using the diagonal approximation we have
\begin{equation}
\label{11}
P(t)=\int_0^\infty \Pi_M(\Gamma_s)\exp(-\Gamma t) {\rm d}\Gamma_s \;.
\end{equation}
For the long-term asymptotic only the narrow resonances are of importance.
Then, substituting \ $\Pi_M(\Gamma_s)$\ by its asymptotic expression
$\Pi_M(\Gamma_s)\sim\Gamma_s^{M-1}$, $\Gamma_s\ll 1$ [see Eq.~(\ref{8})],
we obtain $P(t)\sim(2t/N)^{-M}$. This power law decay takes place
only for a ``coherent'' evolution of the initial state. 
In contrast, the short-term asymptotic of $P(t)$ coincides with an
``incoherent'' evolution, which would take place if one used uncorrelated
matrices $U_{ran}$ in Eq.\,({9}) at each time step. Then, as follows
from the structure of the matrix $U_{ran}$, at each time step the
state vector decreases its norm by the factor $N/(N+M)$ and
\begin{equation}
\label{12}
P(t)=\left(\frac{N}{N+M}\right)^{t}\approx
\exp\left(-\frac{M}{N}t\right) \;.
\end{equation}

\smallskip  
{\bf 4.}
We proceed with the statistics of the resonances for the physical
model (\ref{2}). Numerically we find them as the eigenvalues of the
matrix (\ref{5}), where the parameter $M$ is identical with the 
denominator $q$ in the condition of commensurability (\ref{4}).

\begin{figure}
\begin{center}
\includegraphics[height=8.7cm,angle=-90,clip]{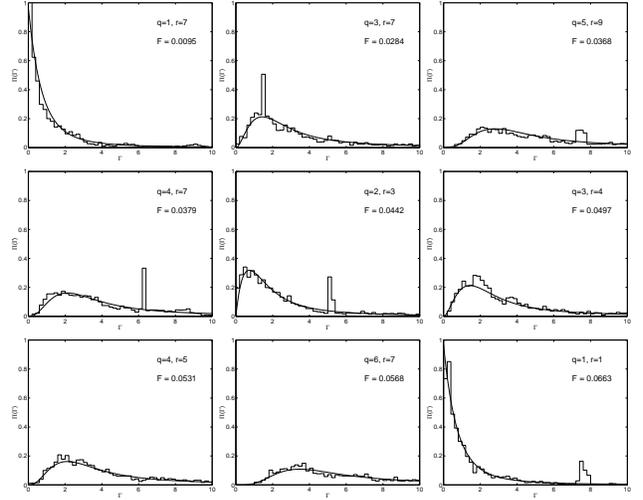}
\end{center}
\caption{Distribution of the scaled resonance width $\Gamma_s$
of the system (\protect\ref{2}) for some values of the static force $F$
satisfying the commensurability condition (\protect\ref{4}). The other  
system parameters are $\omega=10/6$, $\epsilon=1.5$, and $\hbar=0.25$. 
The smooth curves are the distribution (\protect\ref{8}) for $M=q$.}
\label{fig4}
\end{figure}
Figure \ref{fig4} compares the distribution of the scaled resonance widths
$\Gamma_s=\pi\Gamma/\Delta$ in the system (\ref{2}) against the prediction
of RMT given by Eq.~(\ref{8}). It seen that the global features of
the distribution $\Pi(\Gamma_s)$ fit well to the result of RMT.
(The peak-like peculiarities of $\Pi(\Gamma_s)$ are due to the
resonances associated with the stability islands of the classical phase
space. In principle these resonances should be removed from the
analyzed data.) 

We would like to note that to satisfy the condition
(\ref{4}) we adjusted the amplitude of the static force $F$ (the other
system parameters are kept fixed). By changing $F$ we actually change
the classical properties of the system [in particular, the distribution
of the classical delay time (\ref{3})]. Nevertheless, the quantum
distribution $\Pi(\Gamma_s)$ remains practically unchanged
(see cases $F=0.0095$ and $F=0.0663$, for example) and is defined
exclusively by the channel number $q$. This fact clearly demonstrates
the applicability of RMT for the system under study.

We come to the problem of the probability decay. In our numerical
analysis of the system we calculated the dynamics of probability $P(t)$
by two different methods. The first method utilizes Eq.~(\ref{9}) where
the random matrix $U_{ran}$ is substituted by the matrix (\ref{5}).
The second method is the direct numerical simulation of the 
wave-packet dynamics of the system (\ref{2}). The latter method has
the advantage that it allows to study the incommensurate case
but it is essentially more time consuming.
In the commensurate case $T_B/T_\omega=r/q$ (with relatively small
$r$ and $q$) both methods give the same result.
\begin{figure}
\begin{center}
\includegraphics[height=8.0cm,clip]{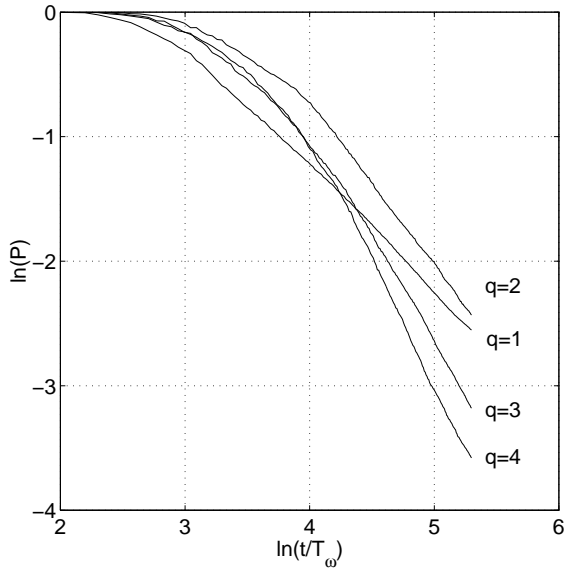}
\end{center}
\caption{Decay of the probability of the system (\protect\ref{2}) for the 
values of $F$ corresponding to $r/q=1$, $3/2$, $4/3$, and
$5/4$ in Eq.~(\protect\ref{4}). The other parameters are the same as
in Fig.~4.} \label{fig5}
\end{figure}

Figure \ref{fig5} shows the behavior of $P(t)$ on a double
logarithmic scale for $r/q=1$, $3/2$, $4/3$,
and $5/4$. It is seen that the survival probability follows 
asymptotically a power law $P(t)\sim t^{-\alpha}$. However, the value 
of $\alpha$ ($\alpha\approx1$, $4/3$, $5/3$, $2$, respectively)
differs from that predicted by Eq.~(\ref{10}). 
This means that for very small resonance width (not resolved on
the scale of Fig.~\ref{fig4}) the actual distribution 
$\Pi(\Gamma_s)$ deviates from the distribution (\ref{8}).
For the moment we have no explanation for this deviation from RMT.

It is also worth to note that the algebraic decay discussed is
actually a transient phenomenon for physical systems. The point
is that RMT deals with an infinite ensemble while in practice it is always
finite (and often consists of a single representative).
For a finite ensemble most narrow resonance exists and, thus, a very
far asymptotic is again an exponential decay with the increment
given by the width of this most narrow resonance.


%
\end{document}